# Spontaneous nonreciprocal transport in a gate-tunable ferromagnetic Rashba 2-dimensional electron gas

Gabriel Lazrak[1], Radu Abrudan[2], Börge Göbel[3], David Hrabovsky[4], Chen Luo[2], Victor Ukleev[2], Srijani Mallik[1,5], Luis M. Vicente-Arche[1], Florin Radu[2], Sergio Valencia[2], Annika Johansson[6*], Agnès Barthélémy[1] and Manuel Bibes[1*]

[1] Laboratoire Albert Fert, CNRS, Thales, Université Paris-Saclay, 91797 Palaiseau, France

[2] Helmholtz-Zentrum Berlin, Albert-Einstein-Strasse 15, 10405 Berlin, Germany

[3] Institut für Physik, Martin-Luther-Universität, Halle-Wittenberg, 06099 Halle (Saale), Germany

[4] Plateforme Mesures Physiques à Basses Températures, Sorbonne Université, Campus Pierre et Marie Curie, 4 place Jussieu, 75005 Paris, France

[5] Saha Institute of Nuclear Physics, 1/AF, Bidhannagar, Kolkata 700 064, India

[6] Max Planck Institute of Microstructure Physics, Weinberg 2, 06120 Halle (Saale), Germany

The broken inversion symmetry at interfaces of complex oxides gives rise to emergent phenomena, including ferromagnetism and Rashba spin-orbit coupling (SOC), which profoundly influence the electronic structure by entangling spin and momentum. While the interplay between Rashba SOC and ferromagnetism is theoretically intriguing, its experimental manifestations remain largely unexplored. Here, we demonstrate that ferromagnetic 2DEGs at $SrTiO_3$-based interfaces exhibit spontaneous nonreciprocal transport – a distinctive hallmark of Rashba ferromagnets – even in the absence of an external magnetic field. This nonreciprocal response, along with clear signatures of ferromagnetism such as anisotropic magnetoresistance and the anomalous Hall effect (AHE), is strongly tunable by gate voltage. Remarkably, the AHE not only varies in amplitude but even reverses sign, reflecting a subtle interplay between Fermi level position and Berry curvature distribution. These results establish $SrTiO_3$ 2DEGs as a model platform for studying Rashba ferromagnetism and demonstrate active control over transport phenomena in time- and inversion-symmetry-broken systems, paving the way for gate-tunable spintronic devices.

* annika.johansson@mpi-halle.mpg.de ; manuel.bibes@cnrs-thales.fr

In simple two-band systems, spin degeneracy can be lifted either by magnetic exchange, as in ferromagnets, or by Rashba spin-orbit coupling (SOC)[1]. In ferromagnets, bands split in energy according to spin, leading to spin polarization and spontaneous magnetization. In contrast, Rashba SOC leads to a spin-dependent band splitting antisymmetric in momentum and locks the spin perpendicular to momentum, but preserves zero net magnetization at equilibrium. While ferromagnets have long been central to spintronics[2] – as robust spin sources, spin detectors, and nonvolatile memory elements – Rashba systems have more recently attracted attention for their ability to interconvert spin and charge currents via the Edelstein[3,4] and inverse Edelstein effects[5], offering magnetization-free alternatives without stray field.

The coexistence of ferromagnetism and Rashba SOC in a single material system is symmetry-allowed but has rarely been realized[6], despite the rich physics anticipated from the simultaneous breaking of time-reversal and inversion symmetries[7]. Only a few systems – such as heterostructures of magnetic and non-magnetic topological insulators or magnetically doped Rashba semiconductors – have exhibited key signatures of this regime, notably a magnetic hysteresis in nonreciprocal longitudinal transport[8–10], where the electrical resistance differs depending on the direction of current flow due to the combined effects of broken inversion symmetry and magnetism.

Oxide interfaces based on $SrTiO_3$ (STO) or $KTaO_3$ offer a versatile platform for studying Rashba SOC, which is strong and highly tunable via electrostatic gating[11–14]. These two-dimensional electron gases (2DEGs) are also known for gate-controlled superconductivity[15–17], but they are typically nonmagnetic. Magnetism can be introduced in engineered heterostructures[18,19], as shown by Stornaiuolo et al. who inserted an ultrathin magnetic $EuTiO_3$ spacer layer in $LaAlO_3$/STO interfaces, with ferromagnetism evidenced via X-ray magnetic circular dichroism (XMCD) and the anomalous Hall effect (AHE)[18].

Here, we realize a ferromagnetic Rashba 2DEG by combining ferromagnetic EuO with STO. The resulting interface exhibits high-mobility transport at low temperature and robust signatures of ferromagnetism, including magnetic hysteresis loops measured by X-ray Magnetic Resonant Scattering (XRMS), hysteretic anisotropic magnetoresistance (AMR), and AHE. Crucially, the nonlinear magnetoresistance also displays magnetic hysteresis, revealing a spontaneous nonreciprocal response – a hallmark of Rashba ferromagnets – emerging even in zero applied field. Gate tuning allows us to strongly modulate all these transport signatures, and most strikingly, to reverse the sign of the AHE. Our calculations attribute this behavior to a highly energy-dependent Berry curvature arising from the complex multiorbital band structure of STO 2DEGs. These findings establish a model platform for Rashba ferromagnetism and demonstrate gate-controlled access to coupled symmetry-breaking phenomena with potential for spintronic applications.

To generate a ferromagnetic Rashba 2DEG we deposited a thin layer of Eu onto a (001)-oriented STO single crystal by sputtering. Analogous to other reactive metals such as Al[20,21], Eu triggers a redox process at the interface[22]: it becomes oxidized while locally reducing the STO, thereby doping it with electrons. We monitored this process *in situ* using X-ray photoelectron spectroscopy, cf. Fig. 1b. The inset presents spectra near the Ti $2p_{3/2}$ level before and after Eu deposition. Initially, the spectrum shows a single peak ascribed to single valence $Ti^{4+}$. After Eu deposition, a strong shoulder at lower binding energy indicates the presence of reduced species, mostly $Ti^{3+}$, confirming the formation of oxygen vacancies in the STO. The main panel shows XPS at the Eu $3d_{5/2}$ level. The fit reveals that almost 100% of the Eu is in the 2+ valence state, corresponding to ferromagnetic EuO, without the presence of nonmagnetic $Eu_2O_3$ ($Eu^{3+}$). Complementary X-ray reflectometry data of Fig. 1a indicate an EuO thickness of approximately 4 nm, capped with a 15 nm $Al_2O_3$ protective layer.

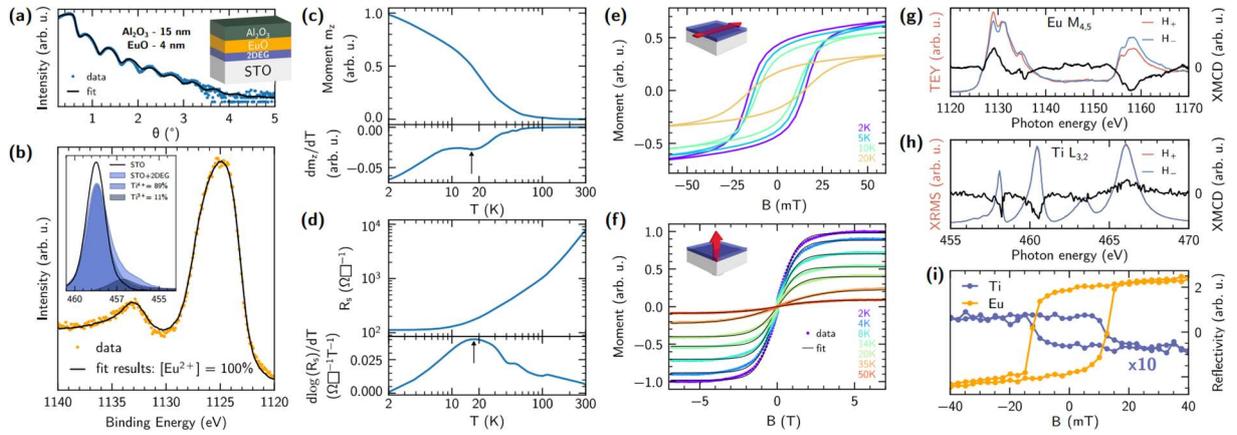

*Fig. 1. Electronic and magnetic properties of EuO/STO ferromagnetic Rashba 2DEGs.* (a) X-ray reflectometry showing EuO and $Al_2O_3$ thicknesses deduced from the fit of the data. (b) XPS spectra: the inset shows Ti $2p_{3/2}$ peaks before and after Eu deposition, revealing $Ti^{3+}$ formation; the main panel displays the Eu $3d_{5/2}$ signal indicating full conversion to $Eu^{2+}$ (EuO). (c) Temperature-dependent resistance (top) and its derivative (bottom), confirming high-mobility 2DEG behavior. (d) Temperature-dependent magnetization (top) and its derivative (bottom), indicating a ferromagnetic transition near 60 K and a secondary feature at ~18 K. (e, f) Hysteresis loops with field applied in-plane (e) and out-of-plane (f), with fits to a $tanh(B/B_C)$ model. (g) XAS and XMCD spectra at Eu $M_{5,4}$ edges at 4 K. (h) Reflectivity and XRMS at Ti $L_{3,2}$ edges at 4 K. (i) XMCD (Eu) and XRMS (Ti) magnetic hysteresis loops.

Fig. 1d shows the temperature-dependent sheet resistance of the EuO/STO sample. The metallic behavior and large residual resistivity ratio (RRR = 67), with no low-temperature upturn, confirm the formation of a high-mobility 2DEG. As expected, the EuO layer does not contribute to charge transport.

The temperature derivative of the resistance (Fig. 1d, bottom panel) shows a broad maximum around 18 K, which – as we will discuss later – coincides with the onset of ferromagnetism in the 2DEG.

Fig. 1c displays the magnetization as a function of temperature. A clear magnetic transition is observed with a Curie temperature ($T_C$) near 60 K, consistent with the ferromagnetic character of the EuO[23]. Interestingly, the derivative of the magnetization also shows an anomaly at ~18 K, hinting at a secondary magnetic feature associated with the interface. Fig. 1e and f presents magnetic hysteresis cycles at different temperatures with the magnetic field applied in plane or out of plane, respectively. The in-plane data confirm that the spontaneous magnetization lies within the film plane. The out-of-plane curves are typical of a hard-axis response in a ferromagnet and can be well reproduced using a phenomenological $tanh(B/B_C)$ model, with $B$ the magnetic field and $B_C$ a characteristic saturation field.

To gain further insight into the magnetism of our EuO/STO samples, we performed X-ray absorption spectroscopy (XAS), XRMS and XMCD across the Eu $M_{5,4}$ and Ti $L_{3,2}$. Europium shows a clear XMCD at $T$= 4 K (Fig. 1g) that agrees with the presence of ferromagnetic $Eu^{2+}$, consistent with the XPS results. A clear XAS signal for Ti, arising from the Ti/Eu interface, was observed. However, the anticipated small XMCD was hindered by the thickness of the Eu layer and the presence of an Al capping. To overcome this limitation, XRMS measurements were performed at 5 degree grazing incidence, see Fig. 1h. A small but clear Ti $L_{3,2}$ edge XRMS was detected. The in-plane field dependence of this dichroic Ti signature shows a hysteresis loop (Fig. 1j) reminiscent of those measured by SQUID (Fig. 1e) and of that obtained at Eu M-edges (Fig. 1j), confirming the FM character of the 2DEG.

Let us now dive into the magnetotransport response of our samples. Fig. 2a shows the low-field magnetoresistance (MR) measured with an in-plane field parallel (top panel) or transverse (bottom) to the current. In both configurations a hysteretic magnetoresistance is observed, with extrema near ±20 mT, matching the coercive field from magnetometry. The opposite signs of the MR in the two geometries are consistent with anisotropic magnetoresistance (AMR), which arises from spin-orbit coupling modifying the electron scattering rate depending on the angle between current and magnetization directions[24].

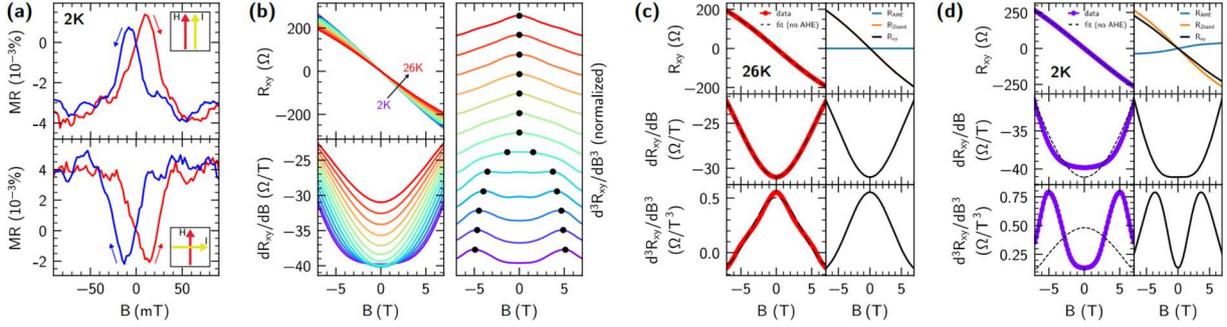

**Fig. 2. Magnetotransport signatures of ferromagnetism in EuO/STO.** (a) Low-field magnetoresistance with magnetic field applied in-plane, either parallel (top) or transverse (bottom) to the current, showing hysteresis consistent with AMR. (b) Hall resistance at various temperatures (top left), along with its first derivative (bottom left) and third derivative (right). A crossover from nonmagnetic (V-shaped) to magnetic (U-shaped and double-peak) behavior occurs below ~14 K. (c, d) Simulations (right panels) of the experimental (left panels) Hall response and derivatives at 26 K (nonmagnetic, no AHE) and 2 K (magnetic, with AHE). Dotted lines in (d) indicate simulations without the AHE term.

Fig. 2b (top left panel) displays the Hall resistance $R_{xy}$ at various temperatures. The curves are clearly nonlinear at low temperature, raising the question of the origin of this nonlinearity. A nonlinear Hall effect is often observed in STO 2DEG even in the absence of magnetism but due to the contributions to the transport response from two types of carriers with different densities $n_1$ and $n_2$ and mobilities $\mu_1$ and $\mu_2$. The Hall resistance then writes:

$$R_{xy}(B) = \frac{-B}{e} \cdot \frac{n_1\mu_1^2 + n_2\mu_2^2 + B^2\mu_1^2\mu_2^2(n_1+n_2)}{(n_1\mu_1+n_2\mu_2)^2 + B^2\mu_1^2\mu_2^2(n_1+n_2)^2} \qquad (1)$$

The transition from linear to nonlinear Hall effect upon electrostatic gating is a way to pinpoint a Lifshitz transition from one-band to two-band transport, that usually occurs for a few $10^{13}$ carriers per cm² (see e.g. Ref. [25]). However, in a magnetic 2DEG, a nonlinear contribution from the anomalous Hall effect (AHE) contribution must also be considered[26].

To disentangle these effects, let us examine the field derivatives of the Hall resistance. The first derivative $dR_{xy}/dB$ (Fig. 2b, bottom left panel) transitions from a V-shaped curve at high temperature (26 K, red trace) – typical of multiband transport in nonmagnetic STO 2DEGs – to a U-shaped curve at low temperature, suggesting the emergence of an AHE caused by ferromagnetism in the 2DEG. This interpretation is reinforced by analyzing the third derivative $d^3R_{xy}/dB^3$, displayed in the right panel of Fig. 2b. At high temperature, the data show a single central maximum; below ~14 K, two symmetric maxima emerge at finite fields $\pm B$, indicating the onset of ferromagnetic behavior.

To quantify this, we perform simulations based on the two-band Hall model. At 26 K (Fig. 2c), the full magnetic field dependence of $R_{xy}$, its first derivative, and its third derivative can be reproduced without any AHE contribution, confirming the absence of magnetism. At 2 K (Fig. 2d), however, the same model fails unless an AHE term is added. Using a phenomenological expression consistent with the magnetization curves of Fig. 1f, we obtain an excellent agreement, capturing both the U-shape in the first derivative and the double maxima in the third.

Fig. 3a shows the temperature dependence of carrier density and mobility extracted from two-band fits to the Hall data. Above 16 K, the fits do not require an anomalous Hall term and yield physically reasonable parameters. Below this temperature, however, the same model begins to fail, yielding unphysical trends in extracted values. To isolate the AHE contribution, we linearly extrapolate the high-temperature values of $n_1, n_2, \mu_1, \mu_2$ down to low temperatures, simulate the ordinary Hall response, and subtract it from the data. This approach yields robust AHE traces for each temperature. Fig. 3b displays the extracted AHE signals as a function of field and temperature. The curves are well captured by a $tanh$-like expression, consistent with magnetization along a hard axis. The AHE amplitude increases sharply below ~14 K, in line with the emergence of ferromagnetism in the 2DEG.

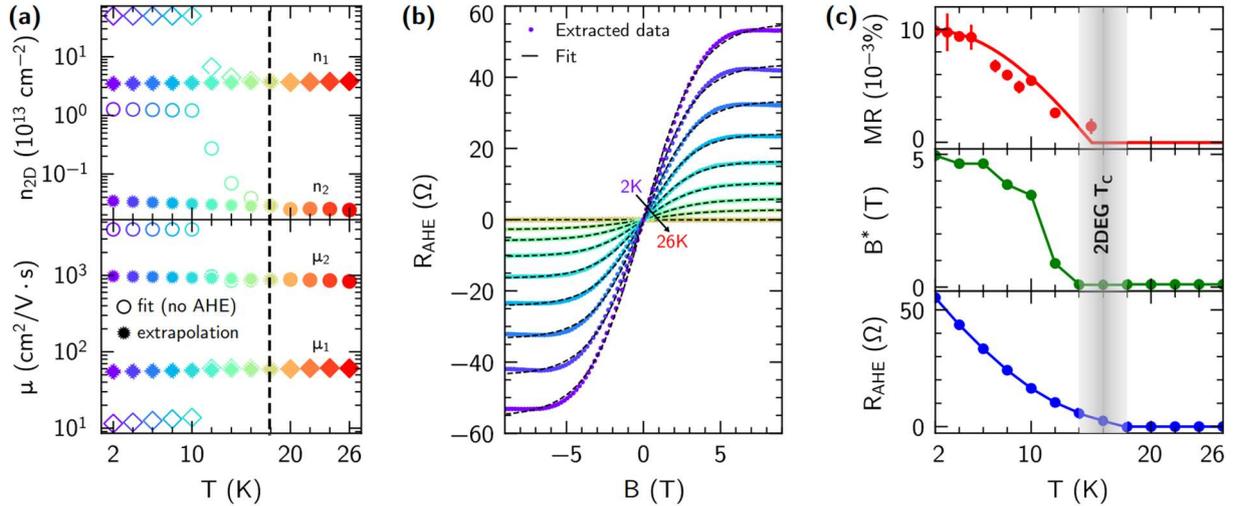

*Fig. 3. Temperature dependence of the magnetotransport parameters.* (a) Temperature dependence of carrier densities $n_1, n_2$ and mobilities $\mu_1, \mu_2$ extracted from two-band Hall fits. Open symbols: direct fits at all temperatures. Solid symbols: extrapolated parameters from fits above 16 K. (b) Anomalous Hall resistance traces obtained by subtracting simulated ordinary Hall response. Fits (solid lines) use a $tanh(B/B_C)$ form. (c) Summary of three magnetic indicators versus temperature: AMR amplitude, field $B^*$, and AHE coefficient $R_{AHE}$, all vanishing near the 2DEG Curie temperature.

Fig. 3c summarizes the temperature dependence of three independent magnetic signatures: the amplitude of the AMR (from Fig. 2a), $B^*$ from $d^3R_{xy}/dB^3$ vs $B$, and the AHE coefficient $R_{AHE}$. All three quantities vanish between 14 and 18 K, defining the Curie temperature of the 2DEG. Notably, this coincides with anomalies in both the resistance and magnetization derivatives (Figs. 1c and 1d). This implies that the temperature dependence of the resistance already shows signatures of the 2DEG magnetic order, and that its magnetization can be detected from the magnetometry data.

Before moving to the gate dependent measurements, let us recall the main features of the band structure of STO-based 2DEGs. It is derived from Ti $t_{2g}$ orbitals and quantum confinement splits these into sub-bands with predominantly $d_{xy}$ or $d_{xz/yz}$ character. The $d_{xy}$ carriers, with lighter effective mass live closer to the interface, whereas $d_{xz/yz}$ electrons are heavier but have longer scattering times due to their spatial extension, and ultimately higher mobility[27]. The $d_{xy}$ bands and the first $d_{xz/yz}$ band have positive and negative Rashba coefficients, respectively, and can hybridize *via* avoided crossings, sometimes resulting in topological band inversions[28]. When combined with exchange splitting from ferromagnetism, the resulting multiorbital band structure becomes highly intricate, with Berry curvature hotspots that evolve strongly with Fermi level position.

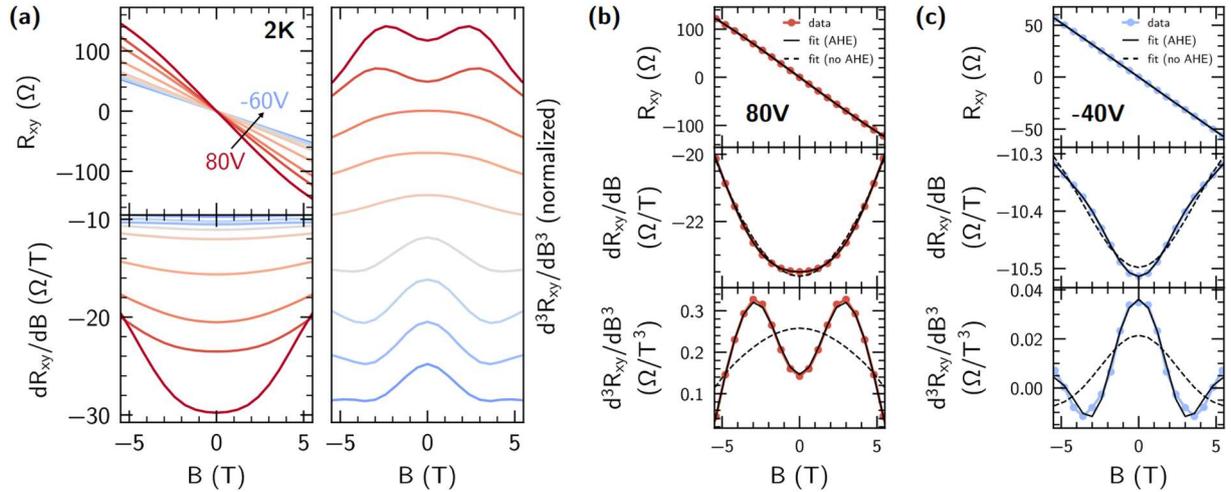

*Fig. 4. Magnetotransport as a function of gate voltage.* (a) Hall resistance for different gate voltages (top left), with corresponding first (bottom left) and third (right) derivatives. Three regimes are identified: AHE-dominated (red), ordinary Hall only (gray), and reversed-AHE-dominated (blue). (b, c) Simulated Hall response and derivatives at 80 V and -40 V, respectively, capturing the AHE sign reversal. All measurements were performed at 2 K.

To explore the influence of the Fermi level position in this rich band structure we applied gate voltages $V_G$ ranging from -60 to +80 V and tracked the evolution of the Hall effect and AMR. Fig. 4a (top left panel) shows $R_{xy}(B)$ at various gate voltages. For large negative $V_G$, the curves are nearly linear, suggesting single-band behavior. At high positive gate voltage, however, nonlinearity emerges and becomes pronounced. This is further evident in the first derivative $dR_{xy}/dB$ (Fig. 4a, bottom left), which evolves from flat to U-shaped, and in the third derivative $d^3R_{xy}/dB^3$ (right), which reveals three distinct regimes. At high positive gate voltages ($V_G$>40 V), the third derivative shows a characteristic double-peak structure, indicating the presence of an AHE. Between +40 V and 0 V, a single broad maximum appears, consistent with nonlinear ordinary Hall response and very small or negligible AHE. At more negative voltages (e.g., -20 V and below), the third derivative displays two minima instead of maxima, signaling a sign reversal of the AHE. To validate this interpretation, we simulate $R_{xy}$, $dR_{xy}/dB$ and $d^3R_{xy}/dB^3$ using the two-band model with a tunable AHE term. Two representative fits at -80V and -40 V are shown in Figs. 4b and 4c, respectively, and closely match the experimental data. The extracted AHE components are summarized in Fig. 5a. Strikingly, the AHE amplitude reverses sign as a function of gate voltage.

In magnetic quantum materials, thin film heterostructure engineering has been used to tune both the amplitude and even the sign of the anomalous Hall effect (AHE). This behavior underscores the intrinsic origin of the AHE, which arises from Berry curvature that can be modulated by confinement, strain, proximity effects, or interfacial symmetry breaking. However, dynamic and reversible tuning of the AHE sign – without changing the carrier polarity – remains extremely rare. To date, it has been observed only in a few cases, most notably in exfoliated MnBi$_2$Te$_4$ flakes[29]. Here, we demonstrate such a sign reversal in a highly versatile oxide thin film system simply fabricated by sputtering at room temperature. Additionally, the AMR shown in Fig. 2a is also gate-tunable, as illustrated in Fig. 5b. Remarkably, its amplitude varies nonmonotonically with $V_G$ and nearly vanishes at large negative voltages, reflecting strong gate control over spin-dependent scattering mechanisms.

Finally, we examine nonreciprocal longitudinal transport, i.e., nonlinear resistance that depends on the direction of the current. In Rashba systems, the direct Edelstein effect gives rise to bilinear magnetoresistance (BMR), which is linear in both electric current and magnetic field[14,30–33]. In ferromagnetic Rashba systems, however, the internal magnetization can itself break time-reversal symmetry, leading to spontaneous nonreciprocity even in zero external field. Figure 5c illustrates this behavior: the nonlinear resistance measured with an in-plane magnetic field transverse to the current shows clear magnetic hysteresis, qualitatively resembling the magnetization loops in Fig. 1e, albeit with an inverted sign. The hysteresis is centered around ±20 mT, consistent with the sample's coercivity. Interestingly, the amplitude of this spontaneous nonreciprocal signal varies strongly with gate voltage and reaches a maximum near $V_G$=-20 V.

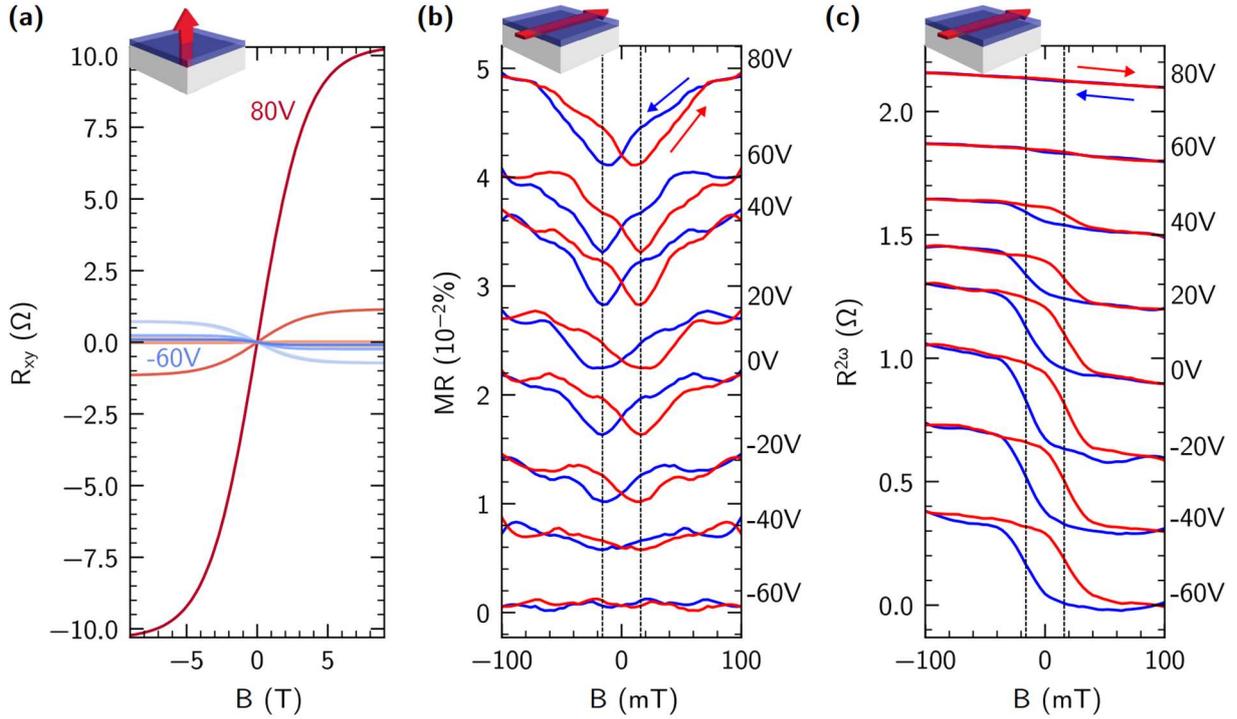

*Fig. 5. Gate dependent AHE, AMR and nonlinear resistance.* (a) Extracted anomalous Hall component as a function of $V_G$, showing a clear sign reversal. (b) Gate dependence of AMR amplitude, peaking at moderate $V_G$ and nearly vanishing at large negative voltage. (c) Nonlinear resistance measured as a function of magnetic field for various $V_G$, showing spontaneous nonreciprocal transport with magnetic hysteresis. All data at 2 K.

We calculate the anomalous Hall conductivity $\sigma_{AHE} = \frac{R_{AHE}}{R_S^2 + R_{AHE}^2}$ and plot its gate dependence in Fig. 6b, together with that of the nonlinear resistance jump at remanence $\Delta R^{2\omega}$ and AMR amplitude $\Delta R^{1\omega}$ in Fig. 6c and d, respectively. Each effect evolves differently with gate voltage – nonmonotonically in most cases. Ferromagnetism thus persists across the entire gating range but manifests through different transport signatures. Focusing on a single observable might therefore give an incomplete or misleading picture.

We now compare the experimental results with calculations. We have built a tight-binding model based on earlier results on STO 2DEGs[13], adding here an exchange term[7], see Fig. 6g. The model consists of four band pairs that are split in energy by ferromagnetic exchange coupling and show avoided crossing due to orbital mixing in the presence of spin-orbit coupling. The exchange coupling (11.6 meV) was chosen to qualitatively reproduce the experimental results. Although a direct experimental determination is not accessible, this value, corresponding to an splitting of the lower bands by 23.2 meV, lies within a reasonable range and is at the lower end of the exchange splitting values estimated in previous studies, such as in Ref. [34]. We compute the carrier density $n_1$ and $n_2$ for bands with dominant $d_{xy}$ and $d_{xz/yz}$ character (respectively) as a function of energy and identify the energy range corresponding to the gate voltage range explored experimentally, focusing in particular on the onset of conduction from the minority $d_{xz/yz}$ carriers, see the comparison between experimental data in Fig. 6a and calculations in Fig. 6e. We then calculated the *k*-resolved Berry curvature distribution for each band, at each energy, integrated it over the full Brillouin zone and thus computed the energy dependence of the intrinsic anomalous Hall conductivity $\sigma_{xy}^{AHE}$ within Kubo transport theory:

$$\sigma_{xy}^{AHE} = \frac{-e^2}{A\hbar} \sum_{k,n} f_{k,n}^0 \Omega_z^n(k) \qquad (2)$$

with $A$ the sample area, $n$ the band index, $f^0$ the equilibrium distribution function and $\Omega_z$ the z component of the Berry curvature. The results of the calculation are displayed in Fig. 6f. $\sigma_{xy}^{AHE}$ is negative at low energy (i.e. low carrier density) and becomes positive near the energy where the minority band becomes populated, as in the experimental data of Fig. 6b. With increasing minority carrier density, $\sigma_{xy}^{AHE}$ strongly increases, again in qualitative agreement with the experimental data. The band-dependent Berry curvature and the band-resolved contributions to the anomalous Hall conductivity are shown in the Supplemental Information, Figs. S1 and S2.

However, quantitatively, the theoretical results differ from the experimentally observed $\sigma_{xy}^{AHE}$ by approximately one order of magnitude. Although the effective four-band model which we use has proven sufficient in describing characteristics of spin-charge interconversion[13], the limited number of bands included might lead to an underestimation of the intrinsic anomalous Hall effect. The Berry curvature can be interpreted as a multiband property that is particularly large in regions of the Brillouin zone where two bands come close to each other. Consequently, considering a limited number of bands might lead to an underestimation of the Berry curvature. Density functional theory calculations[34,35] have shown that $SrTiO_3$-based 2DEGs can be extended over several atomic layers, with multiple subbands intersecting the Fermi level. These additional bands, not considered in our effective model, would modify the band-dependent Berry curvature, and provide additional contributions to the anomalous Hall conductivity. In addition, the quantitative value of $\sigma_{xy}^{AHE}$ strongly depends on the choice of parameters within our model, in particular the inversion asymmetry-induced orbital mixing[13,28] and the exact value of the exchange coupling. Finally, additional extrinsic contributions, which are not considered in our calculations, might contribute to the experimentally observed anomalous Hall signal.

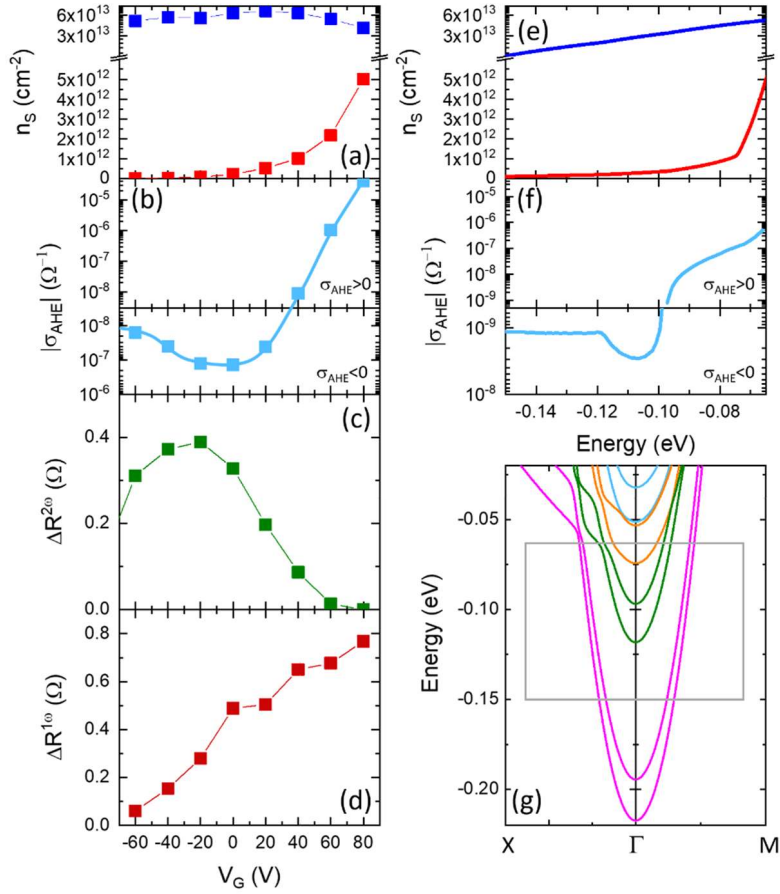

*Fig. 6. Summary of gate-tunable transport signatures and comparison with theory.* (a) Gate dependence of the minority (red) and majority (blue) carriers extracted from the fit of the experimental Hall data. (b) AHE signal, (c) nonlinear resistance jump at remanence and (d) AMR amplitude as functions of gate voltage. Each quantity displays a distinct dependence on $V_G$, reflecting the multiband and Berry-curvature-sensitive nature of the ferromagnetic 2DEG. (e) Calculated energy dependence of the density of carriers with dominant $d_{xy}$ and $d_{xz/yz}$ character. (f) Calculated AHE resistance. (g) Band structure within our tight-binding model used for the calculations.

In summary, we have demonstrated that EuO/STO interfaces host a ferromagnetic two-dimensional electron gas with Rashba spin-orbit coupling, exhibiting gate-tunable magnetotransport signatures including AMR, anomalous Hall effect, and spontaneous nonreciprocal magnetoresistance. Remarkably, the AHE not only varies in amplitude but also reverses sign with gate voltage – an exceptionally rare phenomenon in solid-state systems – arising from the multiorbital band structure and energy-dependent Berry curvature confirmed by tight-binding calculations. The observation of spontaneous nonreciprocity at zero magnetic field provides direct evidence of intrinsic time-reversal symmetry breaking in a Rashba ferromagnet. Previous reports of AHE sign reversal have been largely confined to magnetic topological insulators or ultrathin chiral magnets. In contrast, our work demonstrates this phenomenon in a scalable oxide platform, with an intrinsically generated 2DEG and no need for doping or exfoliation. These results position oxide-based Rashba ferromagnets as a powerful platform to further explore Berry-curvature-driven phenomena and exotic spin textures. The gate-tunable nonreciprocal response, accessible in a simple two-terminal geometry, also suggests new avenues for reconfigurable spintronic logic without reference layers.

**Acknowledgements**

The authors thank Ingrid Mertig, Marc Gabay, Roberta Citro and Mattia Trama for useful discussions. GL and MB acknowledge support for the ERC AdG "FRESCO" Grant No. 833973. This work was funded by the Deutsche Forschungsgemeinschaft (DFG, German Research Foundation) – 545818886 and by the Agence National de Recherche (ANR-DFG project "NOBLESSE").

**Methods**

**Sample preparation.** EuO thin films were grown on (001)-oriented SrTiO₃ (STO) single crystals (5 × 5 mm², Crystec) by magnetron sputtering. Prior to deposition, substrates were sequentially cleaned in acetone and isopropanol using an ultrasonic bath. Hall-bar devices (width W = 20 μm, length L = 200 μm) were patterned by UV lithography. A 4 nm EuO layer was deposited by evaporating Eu metal in an Ar atmosphere onto the STO substrate, followed by deposition of a 15 nm Al₂O₃ capping layer to prevent oxidation. For gate-dependent measurements, a Ti/Au bilayer was sputtered onto the backside of the sample to serve as a back-gate electrode.

**Characterization.** The thickness and interface quality of the films were determined by X-ray reflectivity (XRR) using a Bruker diffractometer. The XRR data were fitted using a multilayer model to extract the EuO and Al₂O₃ layer thicknesses, density, and surface/interface roughness. The chemical composition and valence states of the films were characterized *in situ* using X-ray photoelectron spectroscopy (XPS) on a reference sample grown under identical conditions. After Eu deposition, the sample was transferred to the XPS chamber without exposure to air and measured using Al Kα radiation. Spectra were collected on the Ti $2p_{3/2}$ and Eu $3d_{5/2}$ core levels. Titanium oxidation states were analyzed using CasaXPS software, while europium oxidation states were determined by fitting to reference spectra of metallic Eu, EuO and Eu₂O₃.

**Magnetotransport measurements.** Electrical connections to the Hall-bar devices were established via ultrasonic wedge bonding with Al wires, forming ohmic contacts to the 2DEG. The devices were mounted on chip carriers using silver paint applied to the backside of the STO, which also served as the back gate. A d.c. gate voltage ($V_g$) was applied between the gold back electrode and the device to modulate the carrier density electrostatically. Magnetotransport measurements were carried out in a Quantum Design DynaCool cryostat (PPMS). Temperature-dependent transport was measured using standard d.c. techniques. For gate-dependent studies, an a.c. excitation current $I=|I|\sin(2\pi f t)$ with f=121.27Hz and amplitude of 50 μA was used. Lock-in detection enabled simultaneous acquisition of the first-harmonic longitudinal resistance ($R_{1\omega}$) and second-harmonic component ($R_{2\omega}$). Hall and magnetoresistance signals were antisymmetrized and symmetrized with respect to the magnetic field, respectively.

**SQUID magnetometry.** Magnetization measurements were performed using a vibrating sample magnetometer (VSM) module in a Quantum Design DynaCool SQUID magnetometer. A non-patterned EuO/STO sample was mounted in a plastic straw, with the magnetic field applied either parallel (in-plane) or perpendicular (out-of-plane) to the film surface. Temperature-dependent magnetization was measured under an applied field of 0.1 T. Background contributions from the sample holder and substrate were subtracted from the raw data.

**Synchrotron measurements.** The measurements were conducted using the ALICE reflectometer situated at beamline PM3 of HZB (BESSY II) in Berlin. The beamline provides relatively high flux and various x-ray polarizations. The sample was positioned in a reflection geometry between the poles of a magnet ensuring a magnetic field parallel to the sample's surface. Temperature was precisely controlled using a Janis cryostat. X-ray absorption spectroscopy (XAS) and X-ray Resonant Magnetic Scattering (XRMS) in a theta/2theta geometry were measured simultaneously at different incident angles for incoming circularly polarized photons with negative helicity. The XAS signal was detected by means of total electron yield guarantying surface sensitivity. The XMCD signal was computed as the difference between two XAS spectra recorded at remanence after saturation for -H and +H, respectively. The XMCD asymmetry in reflection was calculated as the ratio of the difference and the sum of two XRMS spectra recorded for two saturation fields of opposite sign. While a clear XAS signal was observed at both Eu $M_{4,5}$ and Ti $L_{2,3}$-edges, no XMCD was detected for Ti given its expected small XMCD signal hindered by the thickness of the Eu and Al layers deposited on top. The interfacial Ti FM signal was detected by the enhanced magnetic interfacial sensitivity provided by the reflected signal. Magnetic hystheresis loops for both Eu and Ti were recorded by means of XRMS at energies with sizeable dichroism (1125.7 eV and 466.5 eV, respectively) by varying the external magnetic field.

**Transport calculations.** We model the $t_{2g}$ orbitals relevant for the formation of the 2DEG using an effective four-band tight-binding model, as introduced in Ref.[13]. The magnetic exchange interaction Hamiltonian is modeled as $H_{ex} = V_{ex}\hat{M} \cdot \Sigma$, with $V_{ex}$ the exchange coupling (chosen as $V_{ex}$=11.6 meV to reproduce the experimental results, and in good agreement with density functional theory calculations reported in Ref. 30), $\hat{M}$ the magnetization direction and $\Sigma$ the spin operator.

Within the linear Kubo transport theory, the anomalous Hall conductivity is calculated via the sum over the Berry curvature of all occupied states, as expressed in Eq. (2). The same expression is obtained using a semiclassical approach, in which the intrinsic anomalous Hall effect is interpreted as a consequence of the anomalous velocity.